\documentclass[journal=jacsat,manuscript=article]{achemso}

\usepackage[version=3]{mhchem} % Formula subscripts using \ce{}
\usepackage[T1]{fontenc} % Use modern font encodings
\usepackage{float}
\author{Suvronil Datta}
\email{suvronild@iisc.ac.in}
\affiliation{\textit{Department of Instrumentation and Applied Physics, Indian Institute of Science, Bangalore, 560012, India}}
\author{Saisab Bhowmik}
\affiliation{\textit{Department of Instrumentation and Applied Physics, Indian Institute of Science, Bangalore, 560012, India}}
\author{Harsh Varshney}
\affiliation{\textit{Department of Physics, Indian Institute of Technology, Kanpur 208016, India}}
\author{Kenji Watanabe}
\affiliation{\textit{Research Center for Electronic and Optical Materials, National Institute for Materials Science, 1-1 Namiki, Tsukuba 305-0044, Japan}}
\author{Takashi Taniguchi}
\affiliation{\textit{Research Center for Materials Nanoarchitectonics, National Institute for Materials Science,  1-1 Namiki, Tsukuba 305-0044, Japan}}
\author{Amit Agarwal}
\email{amitag@iitk.ac.in}
\affiliation{\textit{Department of Physics, Indian Institute of Technology, Kanpur 208016, India}}
\author{U. Chandni}
\email{chandniu@iisc.ac.in}
\affiliation{\textit{Department of Instrumentation and Applied Physics, Indian Institute of Science, Bangalore, 560012, India}}

\title
{\textbf{Nonlinear electrical transport unveils Fermi surface malleability in a moir\'{e} heterostructure}}

\begin{document}
	%\pagewiselinenumbers
	
	%\begin{tocentry}
	
	%Some journals require a graphical entry for the Table of Contents. This should be laid out ``print ready'' so that the sizing of the text is correct.
	
	%Inside the \texttt{tocentry} environment, the font used is Helvetica 8\,pt, as required by \emph{Journal of the American Chemical Society}.
	
	%The surrounding frame is 9\,cm by 3.5\,cm, which is the maximum permitted for  \emph{Journal of the American Chemical Society} graphical table of content entries. The box will not resize if the content is too big: instead it will overflow the edge of the box.This box and the associated title will always be printed on a separate page at the end of the document.
	
	%\end{tocentry}
	
	\begin{abstract}
                Van Hove singularities enhance many-body interactions and induce collective states of matter ranging from superconductivity to magnetism. In magic-angle twisted bilayer graphene, van Hove singularities appear at low energies and are malleable with density, leading to a sequence of Lifshitz transitions and resets observable in Hall measurements. However, without a magnetic field, linear transport measurements have limited sensitivity to the band's topology. Here, we utilize nonlinear longitudinal and transverse transport measurements to probe these unique features in twisted bilayer graphene at zero magnetic field. We demonstrate that the nonlinear responses, induced by the Berry curvature dipole and extrinsic scattering processes, intricately map the Fermi surface reconstructions at various fillings. Importantly, our experiments highlight the intrinsic connection of these features with the moir\'{e} bands. Beyond corroborating the insights from linear Hall measurements, our findings establish nonlinear transport as a pivotal tool for probing band topology and correlated phenomena.
    
   % Van Hove singularities in the density of states enhance many-body interactions and induce collective states of matter ranging from superconductivity to magnetism. In magic-angle twisted bilayer graphene, van Hove singularities appear at low energies and are malleable with progressive band filling, leading to a sequence of Lifshitz transitions and resets observable in Hall measurements. However, at zero magnetic fields, transport measurements in the linear response regime have limited sensitivity to the band's topology.~Here, we utilize the longitudinal and transverse nonlinear transport measurements to probe these unique features in twisted bilayer graphene at zero magnetic field. We demonstrate that the nonlinear responses, induced by the Berry curvature dipole and extrinsic scattering processes, intricately map the Fermi surface reconstructions at various partial fillings of the band. Importantly, our experiment highlights that these features are intrinsic to the moir\'e bands, and not induced or stabilized by the magnetic field. Additionally, we show the tunability of the Berry curvature dipole and extrinsic scattering process with an out-of-plane electric field near the conduction band edge. Beyond corroborating the insights from linear Hall measurements, our findings establish nonlinear transport as a pivotal tool for probing band topology and correlated phenomena. 
		
	\end{abstract}
	
	\textbf{Keywords:} Twisted bilayer graphene, Nonlinear transport, Fermi surface reconstruction, Berry curvature dipole, asymmetric scattering.\\
	
	%\section{Introduction:}
	
	Malleability is typically associated with the physical properties of metals, which can be reshaped through external perturbations. The flat bands in twisted bilayer graphene (TBG) present an analogous parameter space, allowing for effective tuning of its density of states (DOS) governed by symmetries and electron-electron interactions.~Ideally, the single-particle low-energy dispersion in TBG results in two van Hove singularities (vHSs) around half filling of the conduction and valence bands, respectively~\cite{Cao2018correlated, Cao2018unconventional, doi:10.1073/pnas.1620140114}. However, the electronic ground states around these vHSs can vary significantly, depending on the flatness of the bands and their response to external stimuli such as electromagnetic fields \cite{doi:10.1126/science.aaw3780, lu2019superconductors}, dielectric environments \cite{saito2020independent, stepanov2020untying, arora2020superconductivity, PhysRevB.100.161102}, and temperature variations \cite{saito2021isospin, Ghawri2022}. Consequently, low-field Hall measurements have revealed a complex set of vHSs and associated Lifshitz transitions at several integer and fractional band fillings  \cite{bhowmik2023spin, bhowmik2022broken, he2021symmetry, polshyn2020electrical, shen2020correlated, park2021tunable}. Furthermore, in stark contrast to conventional materials, TBG bands turn malleable as they get filled (or emptied), revealing a set of transitions called `resets' \cite{wu2021chern}, where the measured Hall densities approach zero while the band remains partially filled. It is unclear whether interaction effects induced by time-reversal ($\mathcal{T}$) symmetry breaking drive these band reconstructions in TBG or if the malleability is intrinsic to the band structure. Therefore, an alternate probing technique that can reveal the inherent band topology in the absence of a magnetic field is highly desirable.\\

	In this context, second-order transport arising under $\mathcal{T}$-symmetric conditions has the capability to probe band topology~\cite{PhysRevLett.115.216806, du2021quantum, ma2019observation, kang2019nonlinear, sinha2022berry, doi:10.1126/science.adf1506}. Two primary driving factors, akin to the anomalous Hall effect, come into play: intrinsic sources such as the Berry curvature dipole (BCD)~\cite{PhysRevLett.115.216806, PhysRevLett.103.087206, harsh_prb_2023_quantum, harsh_prb2023_intrinsic}, and extrinsic factors, particularly skew-scattering and side-jump~\cite{du2019disorder, doi:10.1126/sciadv.aay2497}.~Initial experiments have showcased the nonlinear (NL) Hall effect in WTe$_2$ crystals \cite{ma2019observation, kang2019nonlinear, He2021}, where the requisite symmetry-breaking conditions were satisfied, leading to a finite BCD. The extrinsic scattering of chiral Bloch electrons can also contribute to second-order electric responses, introducing a finite longitudinal component alongside the Hall effect \cite{he2022graphene}. While BCD has been studied in moir\'{e} graphene systems~\cite{sinha2022berry, PhysRevLett.131.066301}, the role of extrinsic scattering phenomena remains less understood, primarily due to the challenge of disentangling intrinsic and extrinsic contributions in the transverse NL voltage. Specifically, the longitudinal NL conductivity is independent of BCD, and it offers a probe for investigating the extrinsic scattering mechanism in isolation.  
	%providing crucial insights into charge carrier distributions within the system. 
	Although previous experiments have reported a finite second-order longitudinal response in graphene moir\'{e} superlattices~\cite{he2022graphene, PhysRevLett.129.186801, PhysRevLett.131.066301}, a quantitative analysis is lacking, mainly due to the absence of a scaling law from theoretical calculations.\\ 
	
	%\section{Results and discussion:}
	
	\begin{figure*}[t!]
		\includegraphics[width=1.0\textwidth]{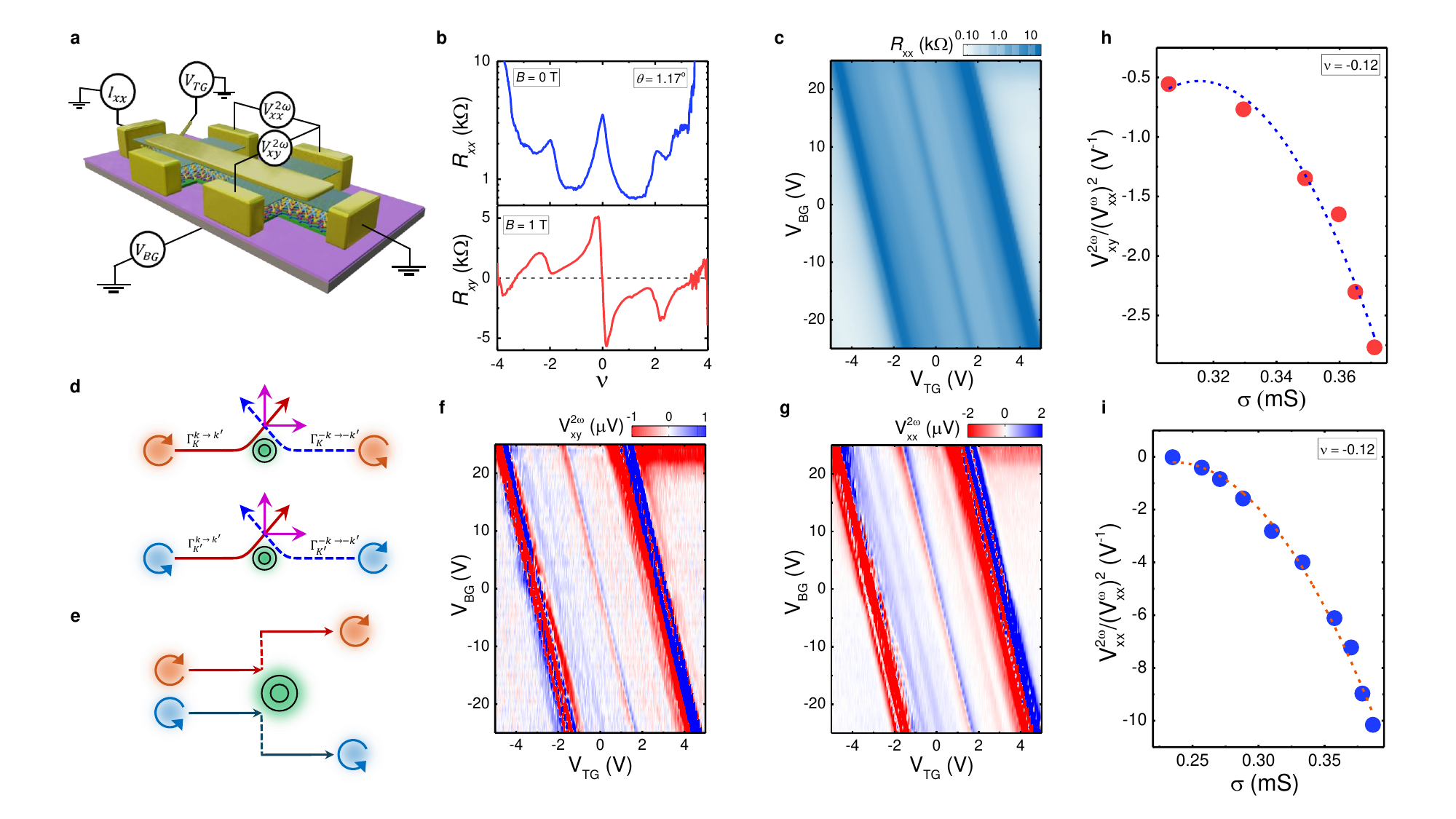}
		\captionsetup{justification=raggedright,singlelinecheck=false}
		{%\textbf{Fig.~1.~Linear and nonlinear transport characteristics}
			\textbf{Fig.~1.~Nonlinear transport in TBG-WSe$_2$ heterostructure:}~~\textbf{a.}~Schematic of hBN-encapsulated TBG-WSe$_2$ heterostructure, overlaid with the measurement scheme. 
			~\textbf{b.}~Four-probe longitudinal resistance $R_{xx}$ (top panel) as a function of band filling $\nu$ measured at a temperature $T=0.3$ K and magnetic field $B = 0$ T and transverse resistance, $R_{xy}$ (bottom panel) measured at $T=0.3$ K and $B = 1$ T. We estimate the twist angle to be $1.17^\circ$, which corroborates well with an alternate estimate from Landau fan diagram.~\textbf{c.}~Colour plot of four probe longitudinal resistance $R_{xx}$ as a function of top gate voltage $V_{TG}$ and bottom gate voltage $V_{BG}$ at $T = 5$~K and $B = 0$~T.~\textbf{d.}~
			%When the electrons have opposite chiralities, 
			The presence of extrinsic scattering in inversion symmetry-broken systems can produce net transverse and longitudinal nonlinear signals. The arrows with red and blue colours indicate different scattering amplitudes. The solid and dashed lines illustrate different directions of electrons, one going from ${k}$ to ${k'}$ and another from ${-k}$ to ${-k'}$, respectively. The top and bottom panels indicate skew-scattering in the two valleys $K$ and $K^\prime$ with opposite chiralities.
			~\textbf{e.}~Schematic of side-jump scattering mechanism. The red and blue colors denote different scattering contributions for opposite chiralities.~\textbf{f-g.}~Colour plots of second order Hall response, $V_{xy}^{2\omega}$ and longitudinal response $V_{xx}^{2\omega}$, respectively as a function of $V_{TG}$ and $V_{BG}$ at $ T= 5$ K and $B = 0$ T, for an excitation current $I^\omega=100$ nA.~\textbf{h-i.}~$V_{xy}^{2\omega}/(V_{xx}^{\omega})^2$ and $V_{xx}^{2\omega}/(V_{xx}^{\omega})^2$ plotted versus linear conductivity, $\sigma$, respectively at $\nu = -0.12$. Dashed lines represent fitting with equation (1) and (2). Finite value of longitudinal nonlinear voltage highlights the presence of extrinsic scattering in the system.}
		\label{Fig1}
	\end{figure*}

In this study, we explore TBG proximitized by tungsten diselenide (WSe$_2$) through NL electrical transport measurements.~Fig.~1a illustrates a schematic representation of our device and measurements. We observe that our dual gated Hall bar device has minimal mixing of the longitudinal and transverse resistivity components (see supplementary information Fig.~S9 and S10), suggesting their independent nature.
 In Fig.~1b, we show the four-probe longitudinal resistance $R_{xx}$ and transverse resistance $R_{xy}$ as a function of band filling $\nu$ at magnetic field $B = 0$~T and $1$~T, respectively (top and bottom panels).~Well-defined resistive peaks in $R_{xx}$ and sign changes in $R_{xy}$ at different $\nu$ are consistent with our twist angle estimation of $\theta \approx 1.17^\circ$ (see Methods and Supplementary Information Fig.~S1).~The negligible variation of $R_{xx}$ at different $\nu$ with a perpendicular electric field arising from dual gate voltages, as shown in Fig.~1c (see Supplementary Information Fig.~S4 for additional set of cantacts), is consistent with prior reports on magic-angle TBG~\cite{doi:10.1126/science.aav1910, Kim2021}.~In order to induce second-order charge current, the breakdown of inversion symmetry is crucial. In our case, ${C}_2$ symmetry of TBG is expected to be broken by the presence of WSe$_2$~\cite{bhowmik2023spin, bhowmik2022broken, doi:10.1126/science.abh2889}.
However, even with broken-${C}_2$ symmetry, the presence of ${C}_3$ symmetry in $\mathcal {T}$-symmetric 2D systems forbids all BCD-induced second-order Hall responses, while the disordered-induced extrinsic contributions (side-jump and skew scattering) can be finite (see Sec. III of the supplementary information). In artificially stacked twisted moir\'e heterostructures, strain is inevitable\cite{Rubio-Verdú2022}. In our TBG sample proximitized by WSe$_2$, it is expected that ${C}_3$ symmetry is broken, making all nonlinear contributions (intrinsic and extrinsic) finite under $\mathcal{T}$-symmetric conditions.
Fig.~1d-e schematically depicts extrinsic scattering (skew-scattering and side-jump) phenomena \cite{doi:10.1126/sciadv.aay2497} assisted by a disorder potential for the two valleys $K$ and $K^\prime$, related by $\mathcal{T}$ symmetry. Here, the transition rate (denoted by $\Gamma^{k\to k'}$ in Fig.~1d) of a Bloch state from momentum ${k}$ to ${k^\prime}$ is not the same as from ${-k}$ to ${-k^\prime}$.~This extrinsic scattering in $K$ and $K^\prime$ valleys with opposite chirality contributes to net voltages in both longitudinal and transverse directions. Fig.~1f-g present color plots of the measured second-order transverse voltage $V_{xy}^{2\omega}$ and longitudinal voltage $V_{xx}^{2\omega}$ as functions of top gate voltage $V_{TG}$ and back gate voltage $V_{BG}$ at $T = 5$ K and $B = 0$~T. Sign reversal is observed at the charge neutrality point (CNP) $\nu=0$, and multiple sign changes take place near-complete band fillings $\nu=\pm 4$ \cite{PhysRevLett.129.186801, PhysRevLett.131.066301}, while peak-like features appear at $\nu=\pm2,\pm3$.\\ 
	
	To understand the origin of different contributions in NL longitudinal and transverse components, we employ a scaling analysis. To this end, we generalize the scaling relationship of transverse response normalized by the quadratic first-order longitudinal voltage $V_{xx}^{\omega}$ defined as $V_{xy}^{2\omega}$/($V_{xx}^{\omega})^2$~\cite{zzdu_Natcom2019_disorder, PhysRevLett.132.026301, huang_arx2023_scaling} with the first-order longitudinal conductivity $\sigma$~(see Supplementary Information Sec.~II). Our theoretical investigation reveals a similar scaling law for $V_{xx}^{2\omega}$/($V_{xx}^{\omega})^2$, with coefficients primarily arising from extrinsic scattering contributions, expressed as (see Supplementary Information for details) 
	
	\begin{eqnarray} 
		\label{eq_1} \frac{V_{xy}^{2\omega}}{(V_{xx}^{\omega})^2} & = &  A_1\sigma^2 + A_2\sigma + A_3~, \\ 
		\label{eq_2} \frac{V_{xx}^{2\omega}}{(V_{xx}^{\omega})^2} & = & A_1^{'}\sigma^2 + A_2^{'}\sigma + A_3^{'}~. 
	\end{eqnarray}
	
The coefficients, $A_{1-2}$ in $V_{xy}^{2\omega}$ and $A_{1-2}^{'}$ in $V_{xx}^{2\omega}$, account for various static (static impurities) and dynamic (phonons) extrinsic scattering contributions, as highlighted in Supplementary Information Sec.~II. The $A_{3}$ term captures the BCD contribution (finite at $T\rightarrow0$) along with dynamic side-jump and skew-scattering contributions, which vanish at $T\rightarrow0$. $A_3^{'}$ originates purely from dynamic extrinsic scattering contributions. We conduct simultaneous measurements of $V_{xy}^{2\omega}$, $V_{xx}^{2\omega}$, and $V_{xx}^{\omega}$ across varying temperatures at different $\nu$. In Fig.~1h-i, $V_{xy}^{2\omega}$/($V_{xx}^{\omega})^2$ and $V_{xx}^{2\omega}$/($V_{xx}^{\omega})^2$ are plotted as a function of $\sigma$ near the CNP ($\nu=-0.12$). Fitting the normalized NL voltages with quadratic polynomials in $\sigma$ validates the presence of both BCD and extrinsic scattering processes in the system.
~Furthermore, we repeat the same measurement for both second-order voltages at various points near the CNP (as shown in Fig.~S5). The fitting remains consistent throughout the whole density range, indicating its robustness.
~A relatively higher magnitude of $V_{xx}^{2\omega}$ compared to $V_{xy}^{2\omega}$ in Fig.~1h-i suggests a significant effect of extrinsic scattering near the CNP. Remarkably, the relative strengths of these coefficients exhibit significant variations upon tuning the carrier density inside flat bands, as discussed later in this article~(see Fig.~3).\\
	
	%------ inserting figure  ----------------%
	\begin{figure*}
	\includegraphics[width=0.8\textwidth]{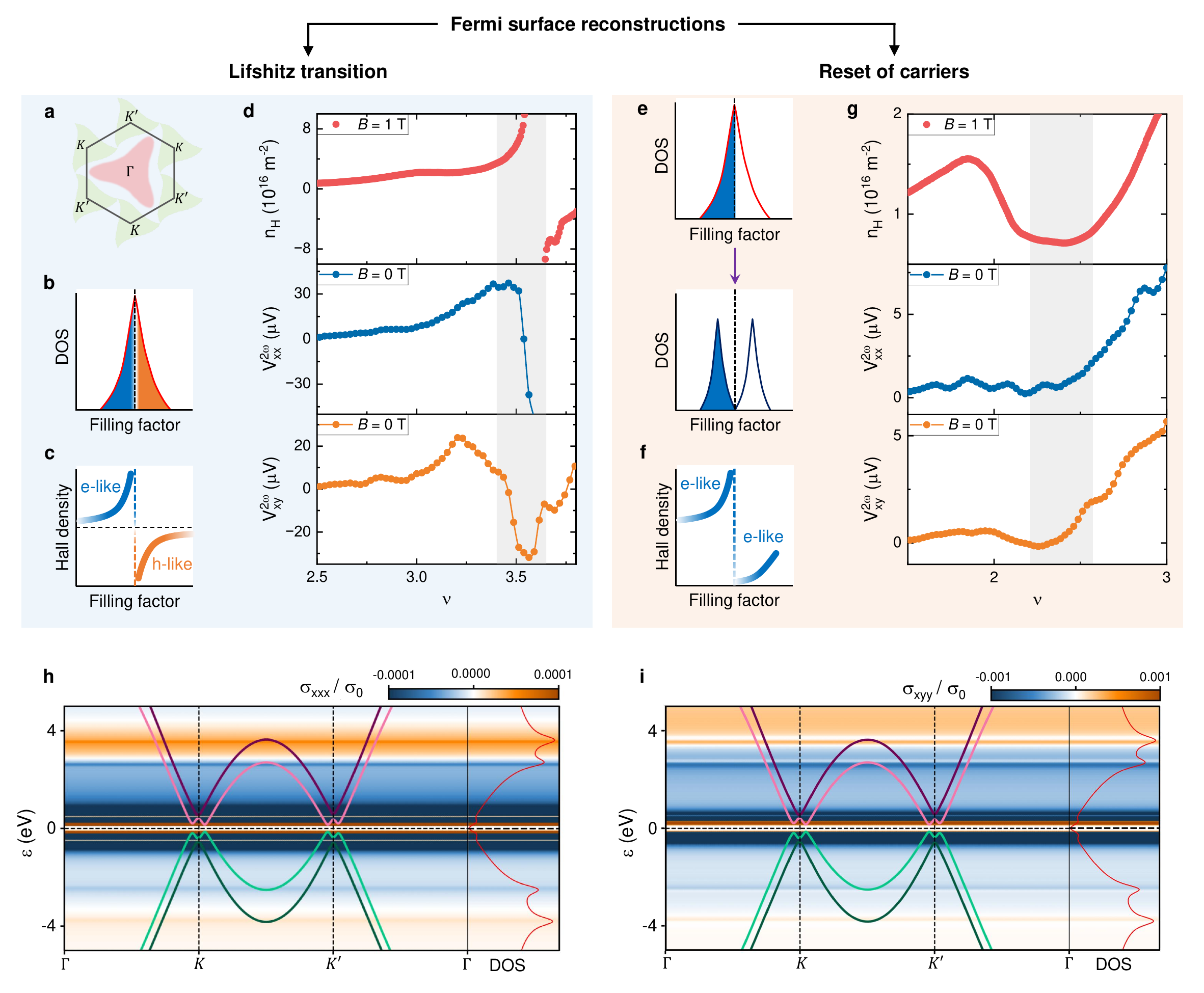}
		\captionsetup{justification=raggedright,singlelinecheck=false}
		{%\textbf{Fig.~2.~Comparison between finite-$B$ linear Hall and zero-$B$ Nonlinear transport data}.
			\textbf{Fig.~2.~Signatures of Lifshitz transition and band resetting in finite-$B$ linear Hall and zero-$B$ nonlinear transport data}.~\textbf{a.}~Fermi contours centered around $K$ and $K^\prime$ (green) and $\Gamma$ (red) points.~As the Fermi energy approaches the van Hove singularity (vHS), the contours around $K$ and $K^\prime$ merge.~At the vHS, the topology of the Fermi surface around $K$ and $K^\prime$ points alters.~\textbf{b.}~Schematic of the DOS vs filling factor ($\nu$) profile, showing a maximum in the DOS. As the Fermi level (dashed line) reaches the middle of the band, an electron-like to hole-like flipping occurs as depicted by a blue-to-orange transition.~\textbf{c.}~Schematic of Lifshitz transition showing a sign reversal in Hall density, $n_H$.~\textbf{d.}~(Top panel) $n_H$ is plotted as a function of $\nu$ adjacent to $\nu = 3.5$ at $T = 5$ K and $B = 1$ T, showing a Lifshitz transition characterized by a sign inversion in $n_H$.~(Middle and bottom panels)~Second harmonic responses, $V_{xx}^{2\omega}$ and $V_{xy}^{2\omega}$ vs. $\nu$ in the same range of $\nu$ at $T = 5$ K and $B = 0$ T.~\textbf{e.}~Schematic representing the `reset' behavior. The DOS splits as the Fermi energy approaches the peak in the DOS (top and middle panel), in contrast to the sign inversion as depicted for Lifshitz transition.~\textbf{f.}~This results in a reset of charge carriers to the same charge carrier type (bottom panel).~\textbf{g.}~(Top panel) The $\nu$ dependence of $n_H$ in the vicinity of $\nu = 2$ at $B = 1$ T and $T = 5$ K. Unlike at $\nu = 3.5$, we find no sign-change in $n_H$. Instead, $n_H$ shows a minimum, followed by a gradual increase.~(Middle and bottom panels)~$V_{xx, xy}^{2\omega}$ plotted as a function of $\nu$ at $B = 0$T showing analogous behavior with a minimum in second-order transport and gradual increase upon increasing carrier density.~\textbf{h-i.}~Colour plots of the calculated nonlinear longitudinal and transverse conductivity for strained bilayer graphene with varying chemical potential ($\mu$). Both colour plots are over the background of the band structure and the DOS. Note the variation of the longitudinal and transverse nonlinear conductivities around the region of vHSs (transition from blue to orange or vice-versa), which qualitatively supports our experimental findings. Here, we have used $\sigma_0 = {e^3 \tau a}/{\hbar^2}$ as the unit of the nonlinear electrical conductivity for 2D systems with $e$ being the magnitude of the electron charge, $\tau$ is the scattering time, and $a$ is the lattice spacing.  
			\label{Fig2}}
	\end{figure*}

	NL transport measurements in mesoscopic samples provide essential information about their physical and band geometric properties.~Early investigations on NL Hall effects in non-centrosymmetric WTe$_2$ crystal \cite{ma2019observation, kang2019nonlinear} highlighted the role of BCD tunable with electric field.
	%induced by the ${\m C}_3$ symmetry breaking. 
	In contrast, studies on graphene/hBN moir\'{e} superlattices have demonstrated extrinsic scattering-induced NL transport \cite{he2022graphene}. Subsequently, topological phase transitions in strained twisted double bilayer graphene were captured through sign-reversal of BCD across the phase transition \cite{sinha2022berry, atasi_2dmat_2022_nonlinear, zhong2023effective}.  
	%establishing NLH as a sensitive tool for probing.
	Recent reports on NL response in TBG suggest both extrinsic and intrinsic origins \cite{PhysRevLett.129.186801, PhysRevLett.131.066301}. 
	However, a detailed understanding of which contribution becomes significant in different parameter regimes is still lacking. 
	%(in terms of the electric field induced bandgap and filling fraction) 
	%where the different contributions play an important role is still lacking. 
	%, although a profound understanding is still limited, primarily due to theoretical constraints in extracting various scattering contributions. 
	Our work attempts to bridge this gap by analyzing NL transport across a wide range of $\nu$.
	%~\textcolor{red}{In contrast to previous reports where NL transport properties were primarily observed at the CNP or band edges \cite{PhysRevLett.129.186801, PhysRevLett.131.066301, atasi_2dmat_2022_nonlinear}, we find finite NL response in a wide range of $\nu$.}
	%These plots capture the correlations and the unique band structure of TBG.} 
To qualitatively understand the data, we have taken a unique approach not reported previously (Table~2 in Supplementary Information). We performed first-order Hall measurements in the presence of a low $B$-field to understand the charge carrier dynamics in the system. The first-order Hall data provides insights into the Fermi surface topology and malleability of the bands possessing vHSs at different integer and fractional $\nu$. Surprisingly, we find a perfect mapping of NL transport to linear Hall data, providing crucial insights into the malleable TBG bands, as discussed in the following section.\\

% NL transport in mesoscopic samples has seen dramatic progress in recent years. While initial studies of NLH on WTe$_2$  \cite{ma2019observation, kang2019nonlinear} emphasized on BCD favored by $C_3$ symmetry breaking, broken $C_2$ symmetry in graphene/hBN moir\'{e} superlattices have shown extrinsic scattering-induced nonlinear transport \cite{he2022graphene}. Subsequently, topological transitions in strained twisted double bilayer graphene were captured through BCD \cite{atasi_2dmat_2022_nonlinear}, establishing NLH as a sensitive probing tool.

Electrically accessible vHSs provide unique opportunities to alter the Fermi surface connectivity and are marked by Lifshitz transitions where abrupt changes of carrier types occur \cite{PhysRevLett.120.096802}.~At the vHSs, the Fermi contours encompassing $K$ and $K^\prime$ points alter in the sense that the winding number drops from $\pm1$ (around $K$, $K^\prime$) to $0$ (around $\Gamma$)~as shown in Fig.~2a. As a result, the measured Hall density $n_H = -(1/e)B/R_{xy}$ where $e$ is the charge of an electron, shows a logarithmic divergence and sign-reversal when the Fermi energy $E_F$ is in the vicinity of the vHS point \cite{Kim2016, wu2021chern, bhowmik2023spin}. This is visually depicted in Fig.~2b-c by a transition from blue (electron-like) to orange (hole-like).~In the case of TBG, where the bands are malleable, the DOS may undergo a different type of phase transition. As the energy bands are successively filled with carriers, the bands may split into two with a small energy gap between filled and empty subbands, as shown schematically in Fig.~2e. Subsequently, $n_H$ abruptly drops to zero or exhibits a minima without a sign change when $E_F$ is inside the gap and increases again upon filling the newly created empty subband (Fig.~2f).~This behavior is called the `reset’ of charge carriers~\cite{bhowmik2023spin, park2021tunable, wu2021chern}~(see Supplementary Information Fig.~S6).\\

In our system, the $\nu$-dependence of $n_H$ reveals a series of Lifshitz transitions and reset of carriers in low $B$-fields (Fig.~2d~and~2g). It remains a question whether such band malleability is 
%driven by broken symmetries and interaction effects 
introduced or stabilized by the finite $B$-field or is an intrinsic property of the system. Our sample shows distinct vHSs and resets at $\nu=\pm3.5$ and $\nu=\pm2$, respectively.~Fig.~2d (top panel) presents $n_H$ as a function of $\nu$ in the vicinity of $\nu = 3.5$ at $B$ = 1~T~(see Supplementary Information Fig.~S7 for the data at $\nu = -3.5$ and $-2$).~The logarithmic divergence of $n_H$ with opposite signs on either side of $\nu = 3.5$ is a clear signature of a Lifshitz transition. We observe a stark resemblance when $V_{xx, xy}^{2\omega}$ are plotted in the same density range but at $B = 0$ T. The middle and bottom panels of Fig.~2d show a clear sign change in $V_{xx}^{2\omega}$ and $V_{xy}^{2\omega}$ both near $\nu = 3.5$. This previously unexplored sign change in the NL Hall voltages across the Lifshitz transition can be understood simply as follows. The NL Hall voltages are proportional to the corresponding NL conductivities (see Eq.~(10) in Sec.~II of the Supplementary Information). Additionally, as shown in Ref.~\cite{zzdu_Natcom2019_disorder} and in Eq.~(3) of Sec.~II of the Supplementary Information, all the second-order NL charge conductivity contributions are proportional to $e^3$. The cubic dependence of all the second-order conductivities on  $e$ indicates that the second-order responses undergo a sign change as the sign of $e$ changes across a Lifshitz transition.%%%%%
~We note that such an $e^3$-dependence is a simple, intuitive approach that explains our data qualitatively. In reality, the observed sign change could be a collective consequence of disparate contributions from BCD, chirality of the electrons and intricate scattering processes (Eq.~(3) of Sec.~II of the Supplementary Information).  Interestingly, other than the CNP and the band gaps, the only density that accommodates a sign change is  $\nu=\pm3.5$, precisely where the Lifshitz transition is realized, closely matching with the Hall density measurements. The observation that such a significant reversal in sign for second-order transport was not seen within the flat bands highlights the role of topology at $\nu=\pm3.5$.\\

We observe a different behavior for the reset around $\nu = 2$. As shown in Fig.~2g (top panel), $n_H$ initially decreases in magnitude near $\nu=2$ (more prominently so near $\nu=-2$, see Fig. S7, Supplementary Information). In contrast to the Lifshitz transition, the sign remains unchanged (similar to Fig.~2f), followed by a sharp increase in $n_H$ when $E_F$ crosses $\nu=2$. 
Remarkably, this malleability of the band structure is also reflected in the second-order transport. We believe that the reduction in the DOS diminishes scattering at $\nu=2$, resulting in a minima in $V_{xx}^{2\omega}$ and $V_{xy}^{2\omega}$ (middle and bottom panels of Fig.~2g). %%%%%%%%
A small negative value in $V_{xy}^{2\omega}$ (lower panel of Fig.~2g) suggests the presence of a finite Berry curvature dipole (as we demonstrate near the band edge in Fig.~4), which does not influence the behaviour of $V_{xx}^{2\omega}$.
%Moreover, relative changes in the magnitude of second-order voltages in the hole side, $\nu=-2$ (as depicted in Fig.~S7), undeniably suggest a closer correspondence with the linear Hall data. Overall, our primary investigation aims to explore phenomena which share similar origins as a first-order response. Unlike linear Hall density, which is primarily governed by the DOS, nonlinear effects are collective phenomena driven by various parameters, where DOS also plays a significant role.}
In summary, the characteristics of $n_H$, $V_{xx}^{2\omega}$, and $V_{xy}^{2\omega}$ align perfectly, highlighting that DOS dependence of scattering and the sign change of the charge carriers across the Lifshitz transition are important contributing factors that govern the observed non-linear effects.~More importantly, two distinct measurement techniques, one under $\mathcal{T}$-symmetric conditions and another with broken-$\mathcal{T}$ symmetry, capture the same features of the system.~These observations confirm that the band reconstructions are intrinsic properties of TBG-WSe$_2$ heterostructure and are not induced by $B$-field.\\

To understand our experimental observations better, we use a simple model of bilayer graphene with a vertical electric field and uniaxial strain and calculate the NL responses. 
The vertical electric field breaks ${\cal C}_2$ symmetry, and strain breaks ${\cal C}_3$ symmetry in our model. We include both BCD and extrinsic scattering contributions in the NL Hall response. In contrast, the longitudinal response originates solely from extrinsic scattering. The details of the calculations are presented in Sec.~II and IV of the Supplementary Information.
In Fig.~2h-i, we display the calculated NL longitudinal~$\sigma_{xxx}$ and transverse conductivity~$\sigma_{xyy}$ for strained bilayer graphene with varying chemical potential~$\mu$ over the background of the band structure and DOS.~Our calculations show that in addition to capturing the changes in band topology near the CNP, NL longitudinal and Hall responses both capture the modulations in the DOS. This can be prominently seen in the energy window between $3-4$~eV on both electron and hole sides, where the double peak structure of the DOS mimics the band resetting-like feature and  Lifshitz transition.\\ 

\begin{figure*}[bth]
	\includegraphics[width=0.8\textwidth]{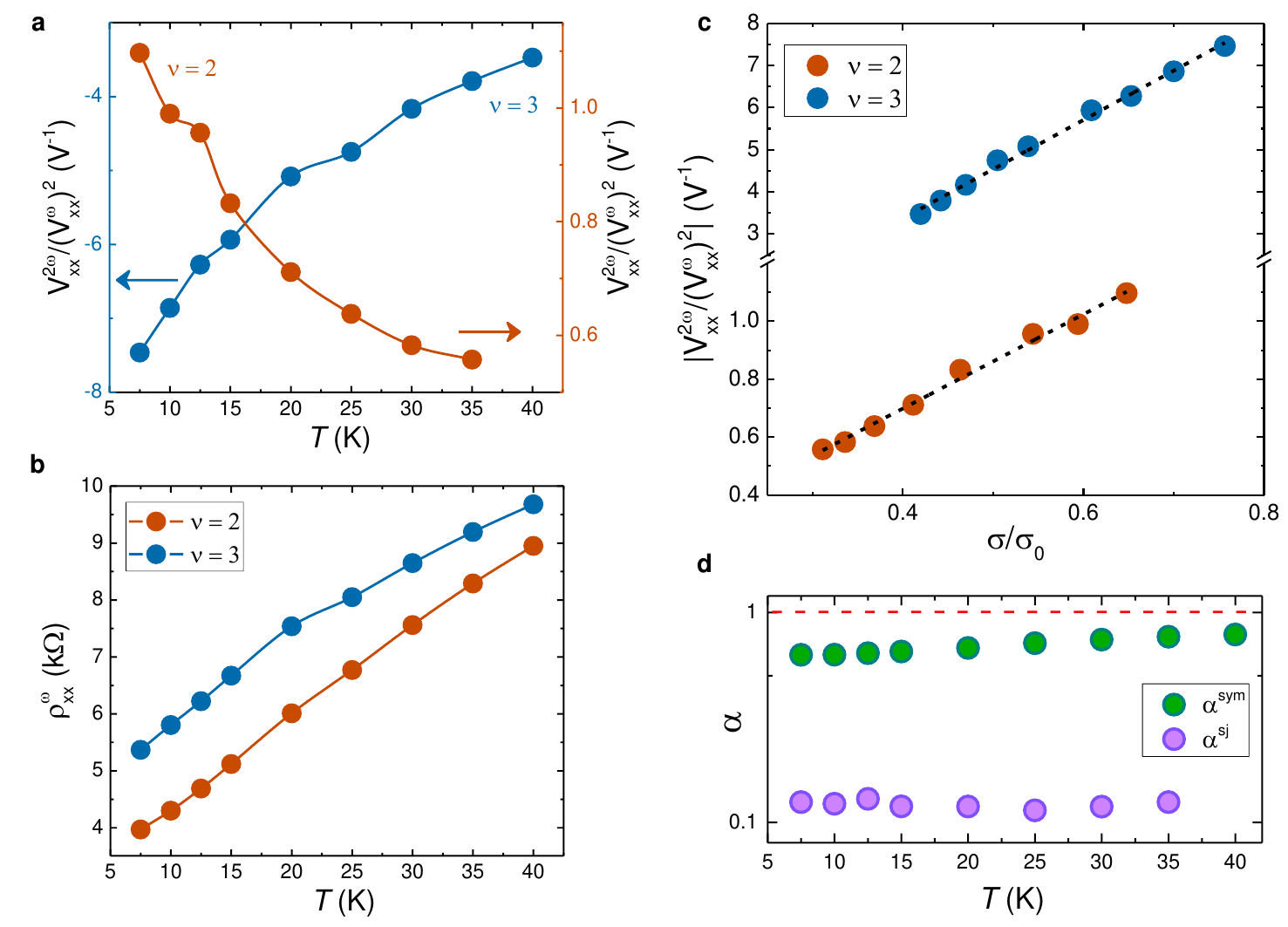}
	\captionsetup{justification=raggedright,singlelinecheck=false}
	{\textbf{Fig.~3.~Temperature dependence of second-order longitudinal response at $\nu \approx 2$ and $3$}.~\textbf{a.}~Normalized second-order longitudinal response, $V_{xx}^{2\omega}/(V_{xx}^{\omega})^2$ as a function of temperature $T$ at $\nu \approx 2$ (orange dots) and $\nu \approx 3$ (blue dots). In both cases, the magnitude reduces as $T$ increases. The solid lines are guide-to-eye.~\textbf{b.}~Temperature dependence of first-order resistivity, $\rho_{xx}^\omega$ measured simultaneously with the second-order transport at, $\nu \approx 2$ (orange dots) and $\nu \approx 3$ (blue dots)}.~\textbf{c.}~$V_{xx}^{2\omega}/(V_{xx}^{\omega})^2$ vs $\sigma/\sigma_0$ as measured from Fig.~3a-b for the same densities, $\nu \approx 2$ (orange dots) and $\nu \approx 3$ (blue dots). Black dashed lines show linear fittings.~\textbf{d.}~$\alpha^{sym} = \frac{1/\tau^{sym}\rvert_{\nu = 2}}{1/\tau^{sym}\rvert_{\nu = 3}}$ and $\alpha^{sj} = \frac{1/\tau^{sj}\rvert_{\nu = 2}}{1/\tau^{sj}\rvert_{\nu = 3}}$ plotted as a function of $T$, extracted from the temperature dependence of $\rho_{xx}^\omega$ and $V_{xx}^{2\omega}/(V_{xx}^{\omega})^2$, where $\tau^{sym}$ is the symmetric scattering time constant estimated from Drude conductivity and $\tau^{sj}$ is defined as side-jump scattering time constant (see Supplementary Information Fig.~S8).
	\label{Fig3}
\end{figure*}

We further analyze our NL longitudinal transport data to gain insights into different scattering contributions at different $\nu$. In Fig.~3a, we present $V_{xx}^{2\omega}/(V_{xx}^{\omega})^2$ %($\propto \rho_{xx}\sigma_{xxx}$ - see Eq.~(10) in SI) 
as a function of $T$ at $\nu=2$ and $3$. The corresponding variation of the linear resistivity $\rho^\omega_{xx}$ with $T$ is presented in Fig.~3b. In order to understand the contribution of side-jump and skew-scattering processes at these fillings, we present the dependence of $V_{xx}^{2\omega}/(V_{xx}^{\omega})^2$ on $\sigma/\sigma_0$ in Fig.~3c. Here, $\sigma_0$ is the residual conductivity estimated by extrapolating $\rho_{xx}^\omega$ versus $T$ data to $T=0$~K. In contrast to observations near the CNP (see Fig.~1i), the variation of $V_{xx}^{2\omega}/(V_{xx}^{\omega})^2$ on    $\sigma/\sigma_0$ at both fillings is linear. Comparing this experimental observation with the scaling in Eq.~(2), we deduce that the coefficient $A'_1$ can be neglected compared to the $A'_2$ at both $\nu$. Furthermore, our detailed analysis (see Eq.~(12) in the Supplementary Information) shows that $A'_1$ arises solely from skew-scattering contributions.~This indicates that for these fillings, the skew-scattering contributions to the NL responses are negligible compared to the side-jump contributions. This can also be inferred from the fact that the linear coefficient $A'_2$, which predominantly captures the side-jump contributions (see Eq.~(13) in the Supplementary Information), dominates the response. Comparing the values of $A'_2$  at $\nu =3$ to $\nu =2$ in Fig.~3c, we find the ratio to be $A'_2(\nu=3)/A'_2(\nu=2) \approx 10$, suggesting a higher value of the side-jump contribution at $\nu = 3$ than at $\nu = 2$.

For a more quantitative comparison, first, we define $\alpha^{sym} = \frac{1/\tau\rvert_{\nu = 2}}{1/\tau\rvert_{\nu = 3}}$, as the ratio of the symmetric scattering rates at $\nu=2$ and $3$, where $\tau$ is the symmetric scattering time constant estimated from Drude conductivity. Similarly, for the side-jump scattering time constant $\tau^{sj}$, we obtain $\alpha^{sj} = \frac{1/\tau^{sj}\rvert_{\nu = 2}}{1/\tau^{sj}\rvert_{\nu = 3}}$ (see Supplementary Information Fig.~S8 for the hole side data).~Fig.~3d displays $\alpha_{sym, sj}$ as a function of $T$. Strikingly, we find that in contrast to $\alpha_{sym}$, the value of $\alpha_{sj}$ remains much smaller than unity for the measured temperature range. This indicates that while the symmetric scattering rates for $\nu=2$ and $\nu = 3$ differ slightly, the corresponding difference in the side-jump scattering rate is much more pronounced. These results suggest that second-order responses are more sensitive to the DOS alterations than the first-order responses.\\ 

\begin{figure*}[bth]
	\includegraphics[width=1\textwidth]{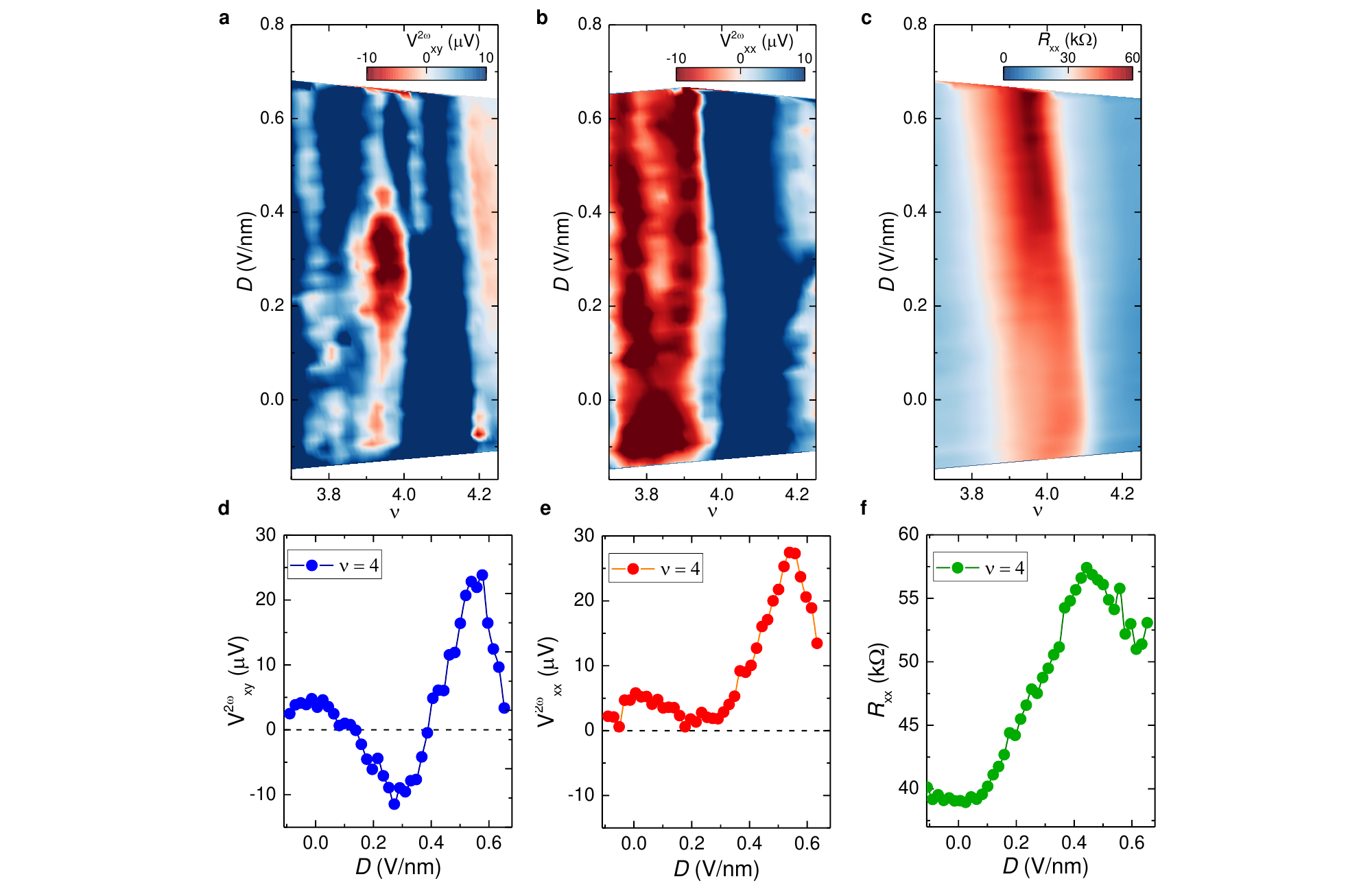}
	\captionsetup{justification=raggedright,singlelinecheck=false}
	{\textbf{Fig.~4.~Displacement field-tunable Berry curvature dipole near $\nu=4$}.~\textbf{a-c.}~Colour plot of $V_{xy}^{2\omega}$,  $V_{xx}^{2\omega}$ and~$R_{xx}$ as a function of $\nu$ and displacement field, $D$ in the vicinity $\nu = 4$.~\textbf{d-f.}~Line plots taken from fig.~4a-c as a function of $D$-field at $\nu = 4$. Though a sign change is seen in $V_{xy}^{2\omega}$, it is absent in $V_{xx}^{2\omega}$.
		\label{Fig4}}
\end{figure*}

In addition to varying scattering strength with temperature, the displacement field $D$ and $\nu$ offer another knob for tuning NL responses. We present the variation of $V_{xy}^{2\omega}$ and $V_{xx}^{2\omega}$ as a function of $\nu$ and $D$-field in the vicinity of the moir\'{e} band edge $\nu\sim4$ (Fig.~4a-b).~In Fig.~4d, we display $V_{xy}^{2\omega}$ as a function of $D$-field at $\nu = 4$, clearly demonstrating two sign changes at $D$-fields of 0.13 V/nm and 0.38 V/nm. While the variation of $V_{xy}^{2\omega}$ with $\nu$ is expected, %as previously observed at integer fillings, 
the striking sign reversals upon changing $D$-fields are noteworthy. Interestingly, no such sign changes with respect to $D$-field are observed in $V_{xx}^{2\omega}$ (Fig.~4b and 4e). The occurrence of sign-reversals exclusively in the transverse response suggests the field-tunable BCD near the moir\'{e} band gap~\cite{PhysRevLett.131.066301}.~The observed behaviour of $R_{xx}$ with $D$-field (Fig.~4c and 4f) indicates the tunability of the first band gap with $D$-field. Overall, the variation of $V^{2\omega}_{xy}$ and $V^{2\omega}_{xx}$ highlight the tunability of the Berry curvature hotspots and extrinsic scattering rate in the TBG flat bands with the $D$-field.\\ %\cite{}.

The investigations of NL longitudinal and transverse transport in TBG showcase their tunability to carrier density, displacement fields, and temperature. Our experiments provide valuable insights by combining conventional first-order Hall and second-order transport measurements to probe the impact of different quantum phenomena on electronic properties and modulations in the DOS. This previously unexplored approach has unveiled multiple Fermi surface reconstructions connected to the vHSs at zero $B$-field. Our study confirms that the $B$-field does not induce or stabilize the probed transitions and highlights their intrinsic nature. The present work establishes NL response as a robust $B$-field-free probe of band malleability and other strongly correlated features manifested in electronic bands. The integration of linear Hall and NL transport measurements offers a promising tool for investigating novel Fermi surface reconstructions and quantum phenomena.

\section*{Supporting Information}
Device fabrication and characterization, measurement scheme, additional experimental results, Fermi surface reconstruction in moir\'{e} materials, scattering time analysis and scaling theory of nonlinear response.
%\bibliographystyle{naturemag}
\bibliography{ref}

\section*{Acknowledgements}
We thank K. Das, S. Das, M. M. Deshmukh, A. Ghosh, E. A. Henriksen, D. Mandal, S. Sinha, and J. Song for helpful discussions and comments. We gratefully acknowledge the usage of the MNCF and NNFC facilities at CeNSE, IISc. H. V. thanks the Ministry of Education, Government of India, for funding through the Prime Minister's Research Fellowship (PMRF). 
U.C. acknowledges funding from SERB via SPG/2020/000164 and WEA/2021/000005. K.W. and T.T. acknowledge support from the JSPS KAKENHI (Grant Numbers 20H00354 and 23H02052) and World Premier International Research Center Initiative (WPI),~MEXT,~Japan.

\section*{Author contributions}
S.D. and S.B. contributed equally to this work.~S.B. fabricated the device.~S.D. and S.B. performed the measurements and analyzed the data.~U.C. advised on experiments. H.V. and A.A. planned all the theory calculations. H.V. executed and analyzed all the theoretical and numerical calculations.~K.W., and T.T. grew the hBN crystals.~S.D., S.B., H.V., A.A., and U.C. wrote the manuscript.~U.C. supervised the project.

\clearpage
\newpage

%\section*{Competing interests}
%The authors declare no competing interests.

\end{document}